# Gas Sensing through Mixed Polarization in Birefringent Porous Silicon Thin Film

Keren Hakshur,[1] Sivan Trajtenberg-Mills,[2] and Shlomo Ruschin[1*]

[1]*Department of Physical Electronics, Faculty of Engineering, Tel-Aviv University, Ramat Aviv Tel-Aviv 69978 Israel*
[2]*Department of Physics, Faculty of Exact Sciences, Tel-Aviv University, Ramat Aviv Tel-Aviv 69978 Israel*
*\*Corresponding author:* ruschin@eng.tau.ac.il

**Abstract:** The interrogation of a sensor based on a highly birefringent film irradiated by an incident beam whose polarization is oriented $45^o$ with respect to the normal film axes is shown to display exceptional features and resolve some outstanding issues in interferometric sensing. The measured pattern as a function of wavelength is complicated displaying a superposition of Fabri-Perot and birefringence oscillations. By means of appropriate optical signal processing based on Fourier analysis, three different periodicity patterns can be elucidated when sensing a single event, improving the sensing process in several ways: it allows sensing at several sensitivity scales, and it clears out the problems of phase and direction ambiguities. A novel effect is observed in which the phase displacement of different patterns move in opposite directions as sensing takes place.

1. **Introduction**

Gas sensing by optical means has attracted much interest in recent decades due to an increasing demand in environment monitoring. Within this context, Porous Silicon (PSi) holds unique advantages as a sensing material as has been reviewed in detail [1]–[3].

In a normal reflection configuration, a simple PSi thin film presents a Fabry-Perot type interference spectrum [4], and its variations, provide a sensitive means of detecting foreign gas infiltration. In more complex configurations, the Fourier Transform (FT) of the optical reflectance spectrum enables a simple tool to compute the effective optical thickness (EOT) of the porous silicon layer [4] and its components. A well-known example of a complex structure where Fourier processing provided useful is that of a multi-layer structure [5]. By a process entitled Reflective Interferometric Fourier Transform Spectroscopy (RIFTS) [6] the FT calculates a spatial frequency spectrum of the reflected light and produces a peak in the spatial frequency whose location along the x axis is related to the EOT product, namely: *2nd*, where *d* is the layer thickness and *n* is the corresponding refractive index. The technique was implemented in a variety of sensing schemes including chemical and bio-sensing [5], providing basically in-depth information of each of the participating layers. A related configuration [7] interrogates simultaneously PSi thin film sections that were placed one beside the other on one plane and demonstrated an optical sensor for ammonia in a humid environment. Light reflection from this arrangement presented a complex interference spectrum, and was resolved by FT, enabling the discrimination between ammonia and water vapor. In a recently published article [8], we reported the implementation of the technique in a highly birefringent film interrogated by light linearly polarized at arbitrary angles. The advantages of sensing based on anisotropic PSi has been previously indicated and demonstrated by several authors [9][10], and high sensitivity parameters were claimed [3]. In the present article we implement the RIFTS technique in conjunction a sensing platform based on [110] oriented PSi sample. Besides the high sensitivity expected, the method is able to resolve several outstanding issues in interferometric sensing.

The interrogation of a Porous Silicon birefringent film by an incident beam whose polarization is oriented $45°$ with respect of the normal crystal axes displays a rather irregular reflection pattern [8]. This is a result of a superposition of three quasi-regular Fabry-Perot fringes corresponding to thin films of thickness $n_e d$, $n_o d$ and $|n_e-n_o|d$. As sensing takes place, changes in the three material parameters namely $\Delta n_e$, $\Delta n_o$ and $\Delta|n_e-n_o|$ can be independently identified by Fourier filtering and are monitored simultaneously in a single sensing event. The reported scheme displays then a unique feature: when fringes are separated by Fourier filtering within a sensing process, the spectral fringes corresponding to the beat $|n_e-n_o|d$, move in *opposite direction* in relation to the single polarization fringes. This relies on the fact that as pores fill with gas or condensed liquid, each of the normal refractive indices increases, while the birefringence as expressed in their difference $|n_e-n_o|$ decreases. This peculiar effect has not been previously reported.

Another aspect is sensitivity: As the displacement of FP fringes for a thin film sensor is inversely proportional to the refractive index, the effective birefringent (BR) pattern depending on the difference $|n_e-n_o|$, will differ significantly in sensitivity as compared to a regular or single-polarization sensor. This sensitivity is reproduced here too by monitoring the peaks corresponding to the ordinary or extraordinary polarization.

A very advantageous feature of the method reported here is that phase and direction ambiguities are removed, as sensing takes place at different FP frequencies and at different scales. These problems are characteristic of single periodicity patterns and limit the dynamic range as dictated by the free-spectral range.

As a last aspect, multi-parametric sensing has obvious implications in overall reliability as the different parameters can be correlated and associated with a single substance under test. All the listed properties were demonstrated by means of a theoretical model and further verified by experimentally collected data.

The reflected spectrum of $45°$ polarized light (with respect to the horizontal axis) has been reported to be anomalous because it lacks a periodic pattern [8]. Therefore spectral changes are explicitly more evident for sensing at mixed polarization (i.e. $45°$) as compared to sensing at a single normal polarization. This is more marked at the beat-cancellation regions. It is a result of the superposition of three FP fringe systems each sensing at different periodicities and different sensitivity scales. The rather irregular pattern recorded at mixed polarization can be clarified by performing a FT on the data.

A phase ambiguity problem may arise in a conventional wavelength interrogation scheme when the sensing process manifests in the shift of interference fringes. If the shift is sufficiently large, fringes in the perturbed interference picture can overlap a different interference order of the reference picture, and if the sensing event is not continuously monitored, this can bring up erroneous quantitative conclusions on the amount of the perturbation. The non-periodical character of a mixed polarization scheme resolves this problem, but this comes at a price of irregularity in the pattern that potentially complicates the quantitative determination of composition changes within the sensor's material. This problem is resolved by multi-scale sensing where the complicated pattern of the dual polarization can be resolved by filtering the peaks in in the FT region as will described further on.

2. **Theory**

A detailed theoretical model which calculates the expected reflection form a birefringent PSi film was described in our previous paper [8]. In the case of sensing, an additional term [V], which accounts for the infiltration of the analyte, (ethanol in our case) vapors into the PSi pores, is included in the Bruggeman model. In addition, the native oxidation, polarization state and depolarization effects are taken into account, resulting in a four component model consisting of silicon, silicon dioxide, pores (air) and the analyte (ethanol) vapors [9] [11]. Accounts for the detailed modeling along with experimental data are presented later-on. In order to give a basic explanation of the effect of sensing at two scales, we provide a simplified model of 4-wave interference that is analytically tractable. In this model, a first polarizer, at $45^0$ divides equally the incoming field $E_0$ into two orthogonal polarizations $H$ and $V$. Each polarization undergoes two reflections at the upper and lower plane of the film respectively (multiple reflections are neglected). The complex amplitudes at the exit of the upper plane are correspondingly: $E_H r_H^{upp}, E_V r_V^{upp}$ for the upper-reflected parts and for the lower-reflected parts: $E_H r_H^{low} \exp(ikn_o 2d), E_V r_V^{low} \exp(ikn_e 2d)$.

Each of these components, upon passing the polarizer for a second time, is projected into the polarizer axis giving a total sum of:

$$E^{TOT} = \frac{1}{\sqrt{2}} (E_H r_H^{low} \exp(ikn_o 2d) + E_V r_V^{low} \exp(ikn_e 2d) + E_H r_H^{upp} + E_V r_V^{upp}) \quad (1)$$

The reflection coefficients for the two polarizations $r_{H,V}^{low,upp}$ be different in general [12]. Expanding the terms and separating them spectrally:

$$|E^{TOT}|^2 = \frac{1}{2} E_0^2 (A_{DC} + A_o \cos(kn_o 2d) + A_e \cos(kn_e 2d) + A_{BR} \cos(k2d(n_o - n_e)))$$

where  (2)

$A_{DC} = (r_H^{upp} + r_V^{upp})^2 + (r_H^{low})^2 + (r_V^{low})^2$
$A_e = 2r_V^{low}(r_H^{upp} + r_V^{upp})$
$A_o = 2r_H^{low}(r_H^{upp} + r_V^{upp})$
$A_{BR} = 2r_H^{low} r_V^{low}$

In this last equation, terms are spectrally separated namely:

1. DC term that does not shift with sensing. The height of the peak depends on the reflection coefficients which depend weakly on the refractive indices

2. Cosine terms with frequencies $(kn_o 2d)$, $(kn_e 2d)$, that will shift according to changes in the respective refractive indices while sensing.
3. A low frequency $(k2d(n_o - n_e))$ term originated on birefringence

Performing a Fourier Transform in the *k* coordinate the diagram will look as displayed in Figs. 1 and 3.

The expectation from a standard birefringent PSi sample in gas sensing, is that as the gas fills (condensate) the pores, the two refractive indices $n_e$ and $n_o$ increase but the birefringence $n_e - n_o$ decreases and the broad peak shifts in the opposite direction as compared to the finer modulation peaks as indicated in Fig. 1.

### 3. Experimental

The birefringent PSi monolayer was fabricated by electrochemical etching of a p-type (110) oriented Si wafer with resistivity of 0.003-0.005 Ω·cm. The sample was etched for four minutes at a current density of 50 mA/cm$^2$. The PSi sensor was not thermally oxidized because the oxidation reduces the birefringence of the PSi resulting in decreased sensitivity. We employed a PSi sensor which partially oxidized at room conditions because a freshly processed PSi film has an hydrophobic nature which can diminish or prevent the condensation of analyte inside the sensor pores. Ethanol vapors in $N_2$ gas were produced by bubbling dry $N_2$ through liquid ethanol at room temperature. The ethanol within the nitrogen stream was further diluted by a second nitrogen stream before directing it to the flow cell.

The sensor was placed in a sealed flow cell. Light from a tungsten halogen lamp (Ocean Optics LS-1) is linearly polarized by 45° around the beam-propagation axis using a rotating polarizer, after which it illuminates the PSi sensor. The reflected light from the sample is passed again through the polarizer, and is collected into a spectrometer detector (OceanOptics USB4000). A full description of the PSi sample production and the optical setup were provided in our previous papers [8], [13].

### 4. Results and Discussion

The reflected spectra from the birefringent PSi sample was analyzed by calculating its Fourier transform (FT). The resulting spectrum was filtered through a narrow band pass filter after which the inverse FT (IFT) was calculated. The filters were applied on either low or high frequencies of the FT, in order to characterize the different effects of these frequencies in the reflected light of the ethanol infiltration into the PSi pores.

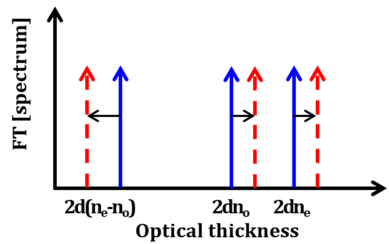

Fig 1. Schematic theoretical shift of the peaks in the Fourier domain. Fourier spectra of the sensor reflectance spectrum with (red dashed lines) and without (blue solid lines) ethanol.

The schematic Fourier spectra of the theoretical shift of the peaks, with (red dashed lines) and without (blue solid lines) ethanol according to the simplified model described above is shown in Fig 1. Here, a clear difference between the high and low frequencies is visible, where the low frequency experiences a shift to the left while the high frequency experiences a shift to the right. The reflection spectra, both theoretical and experimental, at 45° polarization are shown in Fig 2 before and after infiltration of the gas mixture. As seen the main features of the measured spectrum are reproduced by our previously published model [8]. The complicated pattern of the dual polarization seen in Fig. 2, can be resolved into quasi-periodic patterns, by filtering the data around peaks in the FT graph for high and low frequencies.

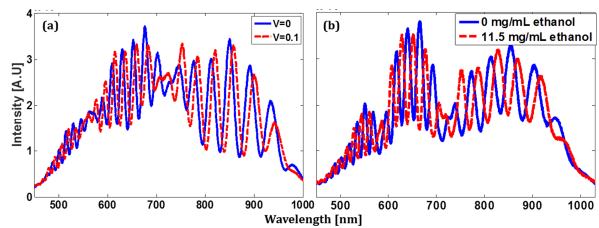

Fig 2. Reflection spectra (a) theoretical (b) experimental where the solid line (blue) presents the PSi reflection and the dash line (red) presents the Psi with ethanol. $V$ is the liquid fraction of the condensed ethanol vapor in the sensor pores.

The simulated and experimental values of the reflected spectra FT can be seen in Fig (3a-b). Here, a clear difference between the high and low frequencies is visible, where the low frequency experiences a shift to the left while the high frequency experiences a shift to the right. Three clear peaks are differentiated corresponding to optical thicknesses of $n_{ord}d$, $n_{ext}d$ and $|n_{ord}-n_{ext}|d$.

The opposite shift is clearly observed for the peaks and confirmed by both the detailed and the simplified model. In Fig. 3 (b), which presents the experimental Fourier spectra of the sensor, the shift of the peak which presents the $|n_{ord}-n_{ext}|d$ is not noticeable due to the limited display resolution.

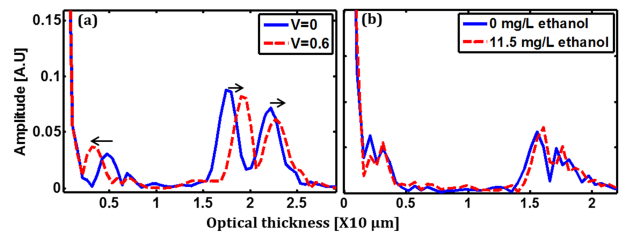

Fig. 3 Fourier spectra of the sensor reflectance spectrum with (red dashed lines) and without (blue solid lines) ethanol. (a) Simulated results corresponding to the theoretical model. (b) Experimental results.

The filtering enables also to observe how the BR sensing solves both, the phase and directional ambiguity resolution: Indeed, looking at Fig. 4(b), which would be equivalent to sensing in a conventionally (non-BR) configuration, the higher concentration trace, moves more than half a period with respect to the reference trace. In a blind sensing scenario, the observer looking at the dashed and dotted traces would be unable to discern whether the concentration of the analyte increased by a large amount or decreased by a smaller one. This ambiguity is clearly resolved by filtering around the low-frequency BR peak of Fig 4(a), which monotonically shifts the peak by a small fraction of π.

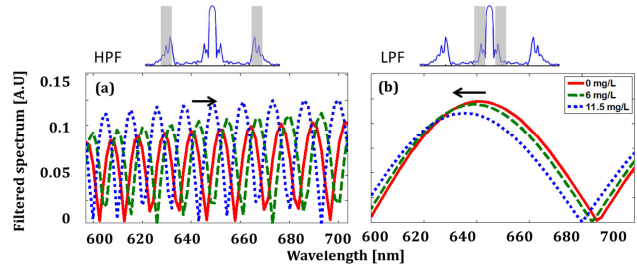

Fig 4: Directional ambiguity. Both images show the filtered spectrum after passing through a (a) high pass filter (HPF) with a center frequency of 15.57 1/μm and spatial width of ~1 1/μm, and (b) low pass filter (LPF) with a center frequency of 2.057 1/μm and width ~1.5 1/μm. Top shows the schematic Fourier spectra, gray areas indicate the filtered locations.

The graphs in Fig. 4 also show clearly shifts at opposite directions in a single sensing process commented before. Sensitivity measurements at two-scales is also demonstrated: at approximately 685 nm the higher frequency measured sensitivity was ~0.4 nm/(mg/L) while the lower one was ~0.17 nm/(mg/L). The wavelength sensitivity is not uniform along the measured spectrum, increasing as expected for longer wavelengths as seen in Fig. 5. In comparison, at approximately 820 nm the higher frequency measured sensitivity was ~0.55 nm/(mg/L), while the lower one was ~0.19 nm/(mg/L).

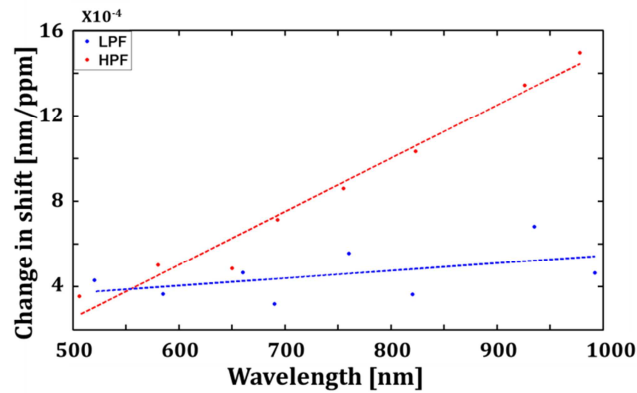

Fig 5: Changes of peak shifts as function of wavelength for the LPF and HPF for ethanol concentrations .Each point represents the sensitivity [nm/ppm] at a specific wavelength and the dash lines represent the points fits.

In order to evaluate the sensitivity of the method, the spectral shift of the filtered spectrum (for both high pass and low pass filters) was extracted for different wavelengths. The resulting sensitivity (e.g. the slope of the shift as a function of ethanol concentration) is presented in figure 5. The known trait [9] where sensitivity increases as a function of the wavelength appears for both filters. In addition, we note that the HPF shows higher sensitivity than the LPF. We note that the special features of the sensing method presented here do not imply a reduction in sensitivity in comparison to previously reported works [14] [15].

### 5. Conclusions and summary

We demonstrated the interrogation of a highly birefringent PSi film sensor which displays unique features in interferometric optical sensing. Processing of optical data, based on Fourier analysis, improved the sensing results and allows sensing at several sensitivity scales. This clears out the problems of phase and direction ambiguities. In addition, a novel effect is observed, in which the phase displacement of different patterns occurs in opposite directions as sensing takes place. By means of Fourier analysis the separation of high sensitivity regimes and low sensitivity regimes was obtained from a single recorded optical spectrum. In addition of enhancing the dynamic range, the demonstrated method revealed apparent advantages in overall reliability as the different parameters can be correlated and associated with a single substance under test.

6. **Acknowledgment**. The authors wish to thank Mrs. Larissa Shpissman for her help in preparing the samples.